# Quantum spin Hall effect in III-V semiconductors at elevated temperatures: advancing topological electronics

Manuel Meyer[1,a)], Jonas Baumbach[1], Sergey Krishtopenko[1,2], Adriana Wolf[1], Monika Emmerling[1], Sebastian Schmid[1], Martin Kamp[3], Benoit Jouault[2], Jean-Baptiste Rodriguez[4], Eric Tournie[4,5], Tobias Müller[6,7], Ronny Thomale[6], Gerald Bastard[1,8], Frederic Teppe[2], Fabian Hartmann[1,b)], Sven Höfling[1]

[1]*Julius-Maximilians-Universität Würzburg, Physikalisches Institut and Würzburg-Dresden Cluster of Excellence ct.qmat, Lehrstuhl für Technische Physik, Am Hubland, 97074 Würzburg, Germany*

[2]*Laboratoire Charles Coulomb (L2C), UMR 5221 CNRS-Université de Montpellier, Montpellier, France*

[3]*Physikalisches Institut and Röntgen Center for Complex Material Systems, 97074 Würzburg, Germany*

[4]*IES, Université de Montpellier, CNRS, F-34000 Montpellier, France*

[5]*Institut Universitaire de France, F-75005 Paris, France*

[6]*Institut für Theoretische Physik und Astrophysik, Universität Würzburg, Würzburg, Germany*

[7]*Department of Physics, University of Zurich, Winterthurerstrasse 190, 8057 Zurich, Switzerland*

[8]*Physics Department, École Normale Supérieure, PSL 24 rue Lhomond, 75005 Paris, France*

**The quantum spin Hall effect (QSHE), a hallmark of topological insulators, enables dissipationless, spin-polarized edge transport and has been predicted in various two-dimensional materials. However, challenges such as limited scalability, low-temperature operation, and the lack of robust electronic transport have hindered practical implementations. Here, we demonstrate the QSHE in an InAs/GaInSb/InAs trilayer quantum well structure operating at elevated temperatures. This platform meets key criteria for device integration, including scalability, reproducibility, and tunability via electric field. When the Fermi level is positioned within the energy gap, we observe quantized resistance values independent of device length and in both local and nonlocal measurement configurations, confirming the QSHE. Helical edge transport remains stable up to $T = 60$ K, with further potential for higher-temperature operation. Our findings establish the InAs/GaInSb system as a promising candidate for integration into next-generation devices harnessing topological functionalities, advancing the development of topological electronics.**

Corresponding authors:
  a) Email: manuel.meyer@uni-wuerzburg.de
  b) Email: fabian.hartmann@uni-wuerzburg.de

# INTRODUCTION

In recent years, topological insulators (TIs) exhibiting the quantum spin Hall effect (QSHE) have attracted considerable attention due to their fascinating electronic properties, characterized by an insulating bulk and gapless helical edge states (*1–4*). These properties make them promising for fundamental research as well as for potential device applications using dissipationless and spin-polarized transport of electrons (*5*). After the theoretical prediction of the QSHE in graphene (*1*) and later in inverted HgTe/CdTe quantum wells (QWs) (*6*), its experimental verification was followed by observing the expected quantized resistance within the band gap (*7*). The nature of the edge states was further confirmed by measurements in nonlocal geometries (*8, 9*). In addition, various other material systems have been proposed to host the QSHE associated with the topological band structure (*10–16*). However, to date, any unambiguous observation of the conductance quantization resulting from the QSHE in other systems has only been observed in monolayer $WTe_2$ (*17*) and InAs/GaSb bilayer quantum wells (BQWs) (*18, 19*). While in $WTe_2$ the QSHE was even robust up to T = 100 K, the complex fabrication of these 2D materials hampers the scalability required for widespread technological development. In HgTe QWs, the limiting factor is the maximal operation temperature, i.e., the QSHE was only observed up to 15 K (*20*). Although band gaps up to 55 meV in HgTe QWs in the TI phase are anticipated via strain engineering (*21*), the pronounced temperature dependence of the band ordering and phase transition to a normal insulating (NI) phase ultimately restricts the QSHE to low temperatures, thereby limiting the use of HgTe QWs in topological electronics applications (*22–24*). This leaves the InAs/GaSb material system as a promising candidate in terms of scalability, reproducibility and electrical tunability of helical edge channel transport at elevated temperatures. It benefits from a rather temperature-independent band ordering indicated through independent transport and terahertz spectroscopy techniques (*25–28*) and the maturity of the growth and processing of III-V semiconductors, including the compatibility with Si-based chips (*29*). Further device functionality in this material system arises from the inherent property of the spatially separated electron and hole gases in the InAs and GaSb layers, respectively. This allows for tunability between trivial and topological band orderings, enabling the phase transition between the TI and NI phases imposed by external electric fields (*30*). While such a phase transition was also observed in Na$_3$Bi, which benefits from large TI and NI gaps (*31, 32*), the required electrical fields are difficult to achieve with conventional gates, and the conductance quantization was not observed. However, the inherently small band gap of a few meV in InAs/GaSb BQWs (*30, 33, 34*) limited the maximum observation temperature of the QSHE to 4 K (*19*). One possibility to increase the band gap in inverted InAs/GaSb-based heterostructures is to replace the GaSb layers with Ga$_{1-x}$In$_x$Sb alloys resulting in a higher overlap of the electron and hole wave functions at the InAs/GaInSb interfaces. This replacement substantially enhances the inverted band gap as it was first found for strained InAs/GaInSb superlattices (*35*). This idea was later applied to InAs/Ga$_{1-x}$In$_x$Sb BQWs (*36–38*), in which the band gap could be increased to 25 meV for $x = 40$. Alternatively, the band gap in InAs/GaSb-based QWs can be increased several times by eliminating the structure inversion asymmetry (SIA) inherent in the BQWs by adding an additional InAs layer (*25*). Thus, by combining these two ideas, the band gap in inverted InAs/GaInSb/InAs trilayer quantum wells (TQWs) is predicted to be as high as 50-60 meV for realistic TQW parameters (*25*). Recent magneto-transport studies on gated Hall bars fabricated from InAs/Ga$_{0.65}$In$_{0.35}$Sb/InAs TQWs revealed band gap values of 45 meV (*25, 39*). These relatively large band-gap values should allow excluding parasitic contributions into the QSHE, such as residual bulk conductivity and trivial edge states (*19, 36, 40–42*) and should enable operations at elevated temperatures.

In this work, we report on the observation of a length-independent quantized resistance value up to elevated temperatures in TIs based on InAs/GaInSb/InAs TQWs with a moderate band gap. The absence of length-dependent resistances within the band gap in both local and nonlocal measurement geometries provide compelling evidence of QSHE, suggesting transport exclusively through helical edge states. For devices with deviating resistances in the gap, the gate-training technique is utilized to obtain a quantized resistance value. Lastly, we investigate the robustness of the QSHE against temperature. We find that the resistance in the band gap remains stable up to $T = 60$ K, with the potential to observe the QSHE at even higher temperatures.

**RESULTS**

**Length-independent conductance quantization**

Fig. 1A schematically shows the band-edge diagram for InAs/Ga$_{0.68}$In$_{0.32}$Sb/InAs TQWs confined by outer AlSb barriers grown on a (001) AlSb buffer layer. The broken-gap alignment at the InAs/(Ga,In)Sb interface leads to the possibility of the band inversion in InAs/(Ga,In)Sb-based QWs, when the first electron-like (E1) subband at zero wavevector $k$ lies below the first hole-like (H1) subband, which is realized at certain thicknesses of InAs and (Ga,In)Sb layers. The interaction between the E1 and H1 subbands at a non-zero wavevector opens a hybridization gap in the BQW, resulting in 2D time-reversal invariant TI state(*15*). In contrast, the presence of the second InAs layer in the TQW eliminates the SIA in the growth direction and the band gap under inversion between the E1 and H1 subbands opens mainly due to the confinement effect at zero $k$, similar to the case of HgTe QWs (*6, 7*). As the band structure calculations show, the inverted band gap in such QW geometry greatly exceeds the band-gap values in InAs/GaInSb BQWs with the same layer materials (*25*). Figure 1B presents the position of electron-like and hole-like subbands at $k = 0$ in the TQW as a function of InAs-layer thickness d$_{InAs}$ for GaInSb-layer thickness of $d_{GaInSb} = 10$ monolayers (MLs), where 1 ML corresponds to the half of a lattice constant in the bulk material. As the wave functions in electron-like subbands are localized in the InAs layers, while they are mostly localized in the GaInSb layers in the hole-like subbands, the symmetric TQW can be interpreted as a "double QW for electrons" with a GaInSb middle barrier, which likewise takes on the role of a "QW for holes" (see Fig. 1A). In this case, the E1 and E2 levels can be interpreted as even-odd state splitting arising from the tunnel-coupled "QWs for electrons". If InAs layers are thin, the E1 subband lies above the hole-like H1 subband, and the TQW has a trivial band ordering, representing a NI state.

Increasing $d_{InAs}$ induces the mutual inversion between E1 and H1 subbands, resulting in inverted band structure of the TQW. The band gap in the inverted TQWs has a nonmonotonic dependence on the InAs layer thickness. As $d_{InAs}$ increases, the gap first becomes indirect, reaching its maximum value when E2 and H1 subbands become close to each other (see Fig. 1B), and then closes due to indirect overlap of the conduction (E2) and valence (E1) bands at a semimetal (SM) state (*25*). A complete phase diagram as a function of $d_{InAs}$ and $d_{GaInSb}$ for symmetric InAs/Ga$_{0.68}$In$_{0.32}$Sb/InAs TQW grown on (001) AlSb buffer is depicted in Fig. 1C. The white hatched region represents the region with the largest TI band gaps above 40 meV, which can be achieved for a given indium content of 32% in the GaInSb alloy. The TQW sample used for the observation of the QSHE consists of two InAs layers with a thickness of $d_{InAs}$ = 30 ML ($\approx$ 9.1 nm) and one Ga$_{0.68}$In$_{0.32}$Sb layer with a thickness of $d_{GaInSb}$ = 10 ML ($\approx$ 3.1 nm). The chosen layer thicknesses result in a topological band structure with a moderate inverted band gap energy of approximately $E_{gap}$ = 27 meV as shown in the 3D dispersion in Fig. 1D. Here, the x- and y-axes represent the [100] and [010] crystallographic directions, respectively. To observe conductance quantization, the lengths between all the contacts of the investigated devices must be smaller than the phase coherence length $\lambda$, which refers to a length scale at which dissipationless edge transport breaks down and counterpropagating spin-up and spin-down channels equilibrate. This results in a deviation of the resistance from the expected quantized value (*43*). Therefore, Hall bars of the investigated InAs/GaInSb/InAs TQW were fabricated, such that contact separation lengths are equal or smaller than what was typically observed in this system (*18, 36, 39*). Details about the sample and band structure calculations can be found in the materials and methods section and in Fig. S1 in the Supplementary Materials. Temperature-dependent measurements to confirm the band gap can be found in Fig. S2 in the Supplementary Materials. Figure 1E shows a typical scanning electron microscopy (SEM) image of such a Hall bar device. The total length of the devices is always $L$ = 10 μm and the width $W$ = 1 μm. All six contacts are symmetrically arranged with a length of $L_C$ = 0.2 μm and are labeled from 0 to 5. For the contact separation length $L_L$ (from middle to middle) between the upper (1 and 2) and lower (4 and 5) contacts, variations of $L_L$ = 1, 2, and 3 µm were chosen. This results in a separation $L_{NL}$ between the outer contacts of $L_{NL}$ = 4.5, 4.0 and 3.5 µm, so L = $L_L$ + 2$L_{NL}$ = 10 µm. Based on the literature (*18, 36, 39*), these lengths should be equal or below the typical phase coherence lengths of TIs based on the InAs/(Ga,In)Sb system. All values for $L_L$ and $L_{NL}$ are colored coded in black, red, and blue. In total, about 20 devices have been studied, where no top-gate leakage current was observed for $V_{TG}$ = +10 to -10 V. Three additional devices were used for each series of measurements reported in Figs. 2, 3 and 4. Therefore, the results in this manuscript are representative for all the devices studied. The results are also reproducible between different thermal cycles.

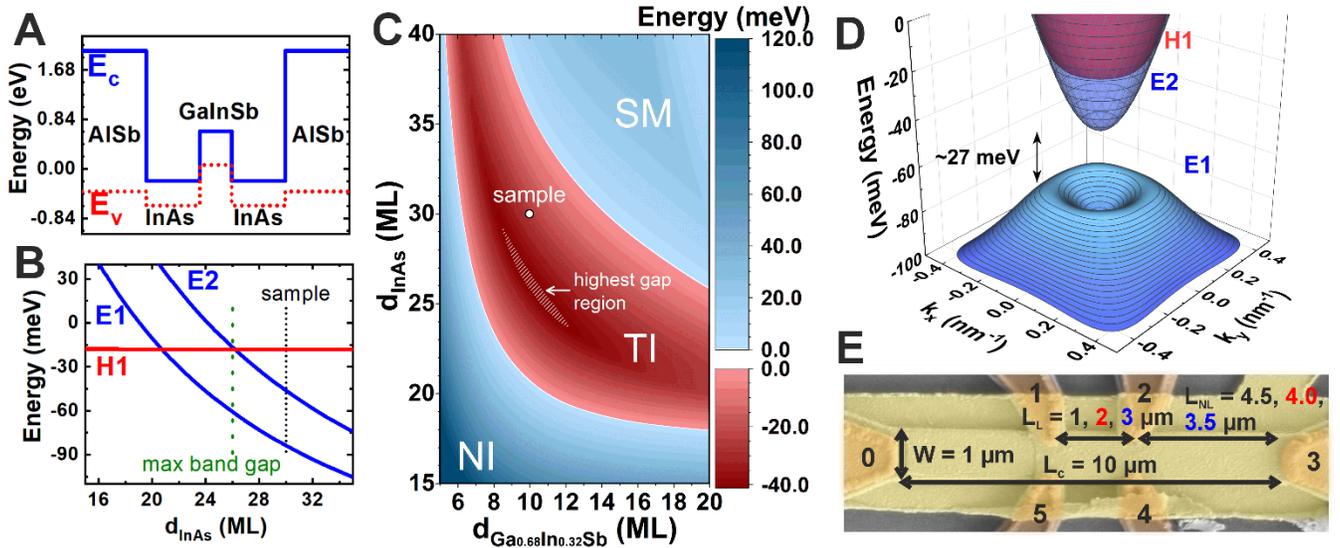

**Fig. 1. Details about the sample.** (**A**) Band-edge diagram for symmetric InAs/Ga$_{0.68}$In$_{0.32}$Sb TQWs grown on (001) AlSb buffer. (**B**) Subband energies at zero quasimomentum *k* in the TQW as a function of InAs layer thickness $d_{InAs}$. Blue and red curves correspond to the electron-like (E1 and E2) and hole-like (H1) states, respectively. The thickness of the GaInSb layer $d_{GaInSb}$ equals 10 monolayers (MLs). (**C**) Color map diagram for symmetric InAs/Ga$_{0.68}$In$_{0.32}$Sb TQWs grown on (001) AlSb buffer as a function of $d_{InAs}$ and $d_{GaInSb}$ with an NI, TI and SM region. The highest band gap region in the TI regime and the sample studied in this manuscript are marked as a white hatched region and with an open symbol, respectively. For the SM region, the band gap is zero, while the energy here marks the overlap between the maximum of the valence band and the minimum of conduction band. (**D**) 3D plot of the band structure of the investigated sample with the x and y axes being the [100] and [010]

crystallographic directions, respectively. (**E**) Colored scanning electron microscopy image of a Hall bar device. The contacts are labeled from 0 to 5 with a length of $L_C = 0.2$ µm and the total length and width of the Hall bar are $L = 10$ µm and $W = 1$ µm, respectively. For the upper and lower contacts, different contact separation lengths $L_L = 1, 2,$ and 3 µm and $L_{NL} = 4.5, 4.0$ and 3.5 µm were chosen to perform length dependent measurements in local and nonlocal geometries.

To observe conductance quantization and prove the QSHE in InAs/GaInSb/InAs TQWs, measurements in both local and nonlocal geometries for different contact separation lengths were performed. If not stated otherwise, all measurements are performed in the dark and at $T = 4.2$ K. The experiments were always conducted in four-probe configuration, which directly eliminates contact resistance. For the typical local configuration shown in Fig. 2A, the top-gate voltage ($V_{TG}$) is swept through the gap, and the resistance $R_{03,12}$ (a 100 nA current is applied from 0 to 3, and the voltage is measured between 1 and 2) is recorded. In the following, all the curves were shifted horizontally to match the position of the peak resistance. For three different lengths $L_L = 1, 2,$ and 3 µm (in black, red, and blue, respectively), the peak resistance value coincides with the expected quantized value of $h/2e^2$ (see Fig. S3 in the Supplementary Materials for a calculation of the expected quantized values). This length-independent resistance in the gap proves the QSHE and ballistic transport through the 1D edge channels as it is expected if the length of the edge channels is smaller than $\lambda$ (*7*). The voltage values of the peak are: $V_{peak} = -2.19$ V, $-1.47$ V and $-1.96$ V for the 1 µm, 2 µm and 3 µm device, respectively. Furthermore, broad resistance plateaus with fluctuations can be observed, as usual for the QSHE at cryogenic temperatures (*7, 44, 45*). The broader peak and larger resistance for the electron and hole regime for the 1 µm device can be explained by a weaker gate efficiency. This is indicated by the weaker dependence of resistance on the top-gate voltage. Therefore, the effective range, where the Fermi energy is tuned in this device, is also smaller than for the other two devices. Transport measurements in the nonlocal configuration (current applied from probes 0 to 1, voltage measured between probes 2 and 3) for different lengths of $L_{NL} = 4.5, 4.0$ and 3.5 µm are presented in Fig.

2B. The voltage values of the peaks for the nonlocal measurements are: $V_{peak}$ = -2.15 V, -2.45V and -2.21V for the 1 μm, 2 μm and 3 μm device, respectively. The different peak positions for local and nonlocal configurations for the same devices result from different gate-voltage ranges, which were used for the experiments. For the same parameters, the peak is always found at the same gate-voltage position irrespective of local or nonlocal geometries, as can be seen in Fig. S4 in the Supplementary Materials. For these devices, a length-independent nonlocal resistance $R_{01,32}$ was observed, which was quantized to the expected value of $h/6e^2$ (41). Additional measurement configurations showing quantization can be found in the Supplementary Materials in Fig. S5. This further confirms the observation of the QSHE and that the transport is provided only at the edges through the helical edge channels in the gap similar to what was previously observed in HgTe/CdTe QWs (8) and InAs/(Ga,In)Sb BQWs (19, 36).

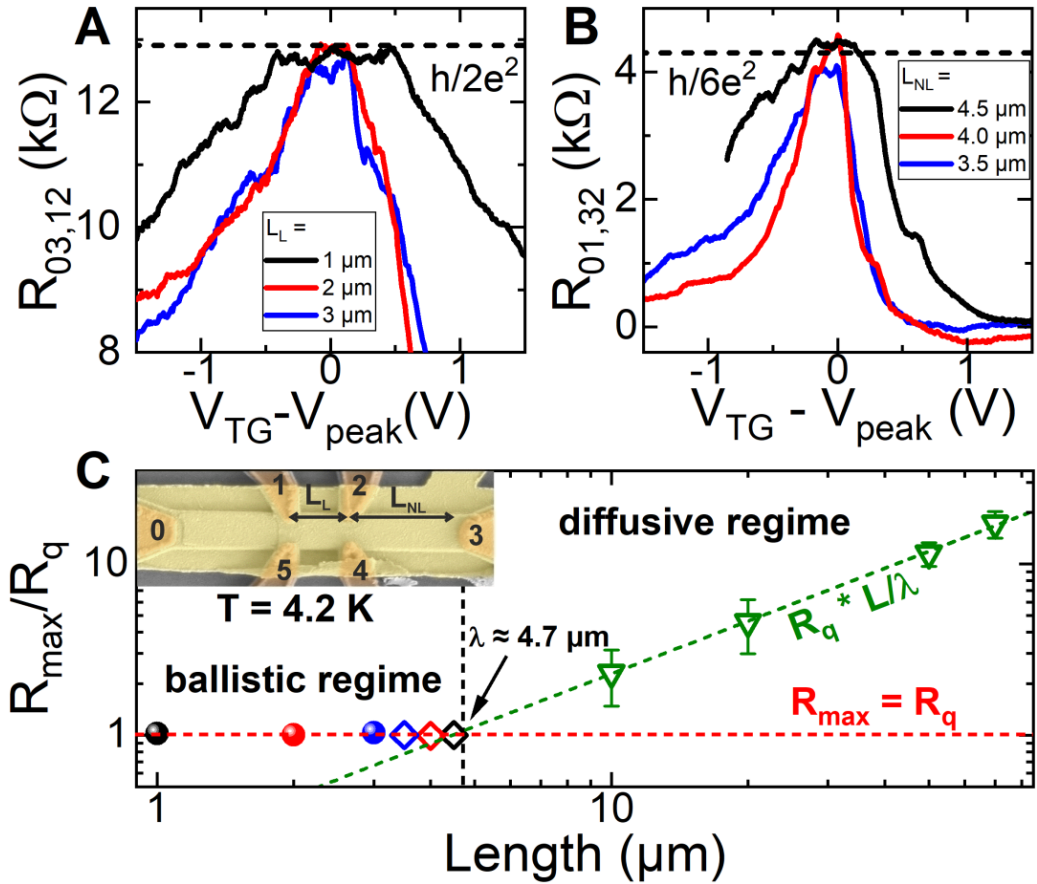

**Fig. 2. Length-independent conductance quantization at 4.2 K.** Local resistance $R_{03,12}$ and nonlocal resistance $R_{01,32}$ (in (**A**) and (**B**), respectively) as a function of $V_{TG}$-$V_{peak}$ for gated devices with different lengths $L_L$ and $L_{NL}$.

The resistance in the band gap is length-independent and quantized to the expected value of $h/2e^2$ and $h/6e^2$, for local and nonlocal configurations, respectively. (**C**) Maximum edge resistance values normalized to the expected quantized value $R_q$ for Hall bar devices of different lengths. Large-scaled Hall bars (green triangles) show a length dependence for the gap resistance in the diffusive regime in comparison to the Hall bars with $L_L$ = 1, 2, and 3 μm in the ballistic regime.

Various devices have been tested to obtain definite results for the length-independence of the resistance. The average values of the maximum resistance $R_{max}$ of all the Hall bars investigated were normalized to the expected quantized value $R_q$ and are summarized in Fig. 2C. $R_{max}/R_q$ = 1 indicates a perfect quantization. The extracted resistances for the devices in both the local (balls) and nonlocal (diamonds) measurement geometries are shown. Up to a length of $L$ = 4.5 μm, the average values are in good agreement with the expected value, which already indicates $\lambda \geq 4.5$ μm. In comparison, the results obtained on large-scaled Hall bars with $W$ = 20 μm and $L_L$ = 10, 20, 50, and 70 μm (green triangles) show a clear length dependence of the edge resistance maximum as expected. These results on the edge resistance for the large-scale devices have been obtained by a detailed analysis to separate edge and bulk contributions (*46*). This demonstrates the transition from the length-dependent diffusive regime to the length-independent ballistic regime as the resistance saturates at the quantized value in comparison to what was observed for trivial InAs/GaSb BQWs (*42*). The quantization in the length-independent regime indicates an insulating bulk and no substantial parasitic contribution to the transport through helical edge channels, as is further confirmed in Fig. S6 in the Supplementary Materials. Furthermore, the intersection of the linear fit with $R_{max}/R_q$ = 1 allows us to extract the phase coherence length to $\lambda \approx 4.7$ μm. A detailed analysis of this extraction can be found in Figs. S7 and S8 in the Supplementary Materials. As all our contact separation lengths in the microscopic devices are smaller than 4.5 μm, this qualitatively confirms the fact that the quantized resistance values are not the result of an arbitrary coincidence but due to the observation of QSHE. The QSHE is further confirmed by a demonstration of breaking of time-reversal symmetry (TRS) in Figs. S9, S10 and S11 in the Supplementary Materials.

**Approaching quantization through gate training**

As previously shown for HgTe QWs (*47*), the phase coherence length of the helical edge states can be increased using the gate-training technique, which is sweeping the gate voltage to minimum values $V_{min}$ and then back through the band gap region. The backscattering and decrease of $\lambda$ is associated to the spin of charged defects in the vicinity of the helical edge states (*46, 48, 49*). In TIs based on InAs/(Ga,In)Sb, such charged defects may also cause backscattering like in HgTe QWs (*50*). The charged defects also lead to a decrease in the quantum scattering time $\tau_q$, which results in pronounced fluctuations in the potential landscape that can be quantified as the quantum level broadening $\Gamma = \hbar/2\tau_q$. This effectively decreases the band gap. By applying the gate-training technique, these charged defects in the vicinity of the TQW are neutralized(*50*). This neutralization improves the quantum scattering time $\tau_q$ and $\Gamma$, and the backscattering is decreased. Therefore, the resistance in the gap approaches quantization for an optimized $V_{min}$ (*47, 50*). The gate-training technique was used on three devices with $L_L$ = 3, 2, and 1 μm from the same TQW, where the resistance deviated from the expected quantized value, as shown in Fig. 3A-C. The top gate was swept to different $V_{min}$ = -2 to -7 V (indicated by a small vertical bar on the left side of all measured curves in Fig. 3A-C) and back through the gap. For the up-sweeps (when sweeping from $V_{min}$ back to positive voltages), the curves are optimized, and the resistances are closest to the quantized value of $h/2e^2$ for $V_{min}$ = -4.5 V (for $L_L$ = 1 and 2 μm) and $V_{min}$ = -3.5 V (for $L_L$ = 3 μm). These curves are marked in bold. By extracting $R_{max}$ and plotting it as a function of $V_{min}$ in Fig. 3D, this improvement can be observed more clearly. The error bars for $R_{max}$ are determined by the resistance fluctuations in the vicinity of the maximum peak values ($R_{max} \pm 5\%$). The inset in Fig. 3D qualitatively illustrates the gate training effect (left: before training, right: after training) on the potential landscape and the neutralization of charged defects in the vicinity of the edge channels. The device with $L_L$ = 3 μm exhibits the largest improvement of the quantization by gate training, whereas only a minimal enhancement was achieved for the device with $L_L = 1$ µm. The lowest values were: $R_{max,3µm} = (1.018 \pm 0.032)$, $R_{max,2µm} = (1.009 \pm 0.038)$, and $R_{max,1µm} = (1.022 \pm 0.017)$, in units of $h/2e^2$. The same minimum resistance value is approached for all devices, which is also the expected quantized resistance value for this configuration. This provides further evidence for the length-independent resistance in the gap and the QSHE, while also highlighting the advantage of gate training in enhancing the reproducibility of quantization.

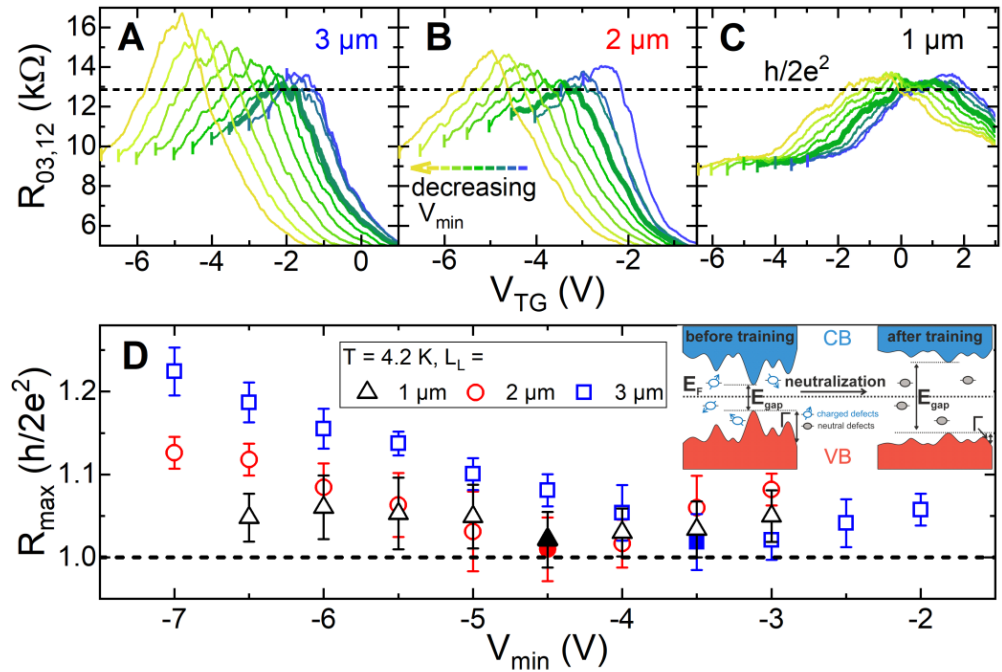

**Fig. 3. Improving the quantization via the gate-training technique.** Gate-training technique was applied to three devices with $L_L = 3$, 2 and 1 µm in (**A**), (**B**), and (**C**), respectively. By sweeping the top gate to an optimized minimum gate voltage $V_{min}$ and back through the band gap, the resistance in the gap is approaching the quantized value of $h/2e^2$ (curves marked in bold). For non-optimized gate voltage sweeps, the coherence length of the edge channels is shorter than the contact separation lengths and hence $R > h/2e^2$. Panel (**D**) shows the evolution of the maximum resistance $R_{max}$ versus $V_{min}$ with the best resistance values close to $h/2e^2$ for optimal gate training shown with filled symbols. The inset qualitatively illustrates the improvement through the gate training (**left**: before training; **right**: after training) on the potential landscape and the neutralization of charged defects in the vicinity of the edge channels.

**Quantum spin Hall effect at elevated temperatures**

As the band ordering in the InAs/(Ga,In)Sb material system is rather temperature insensitive (*26, 27*), we investigated the robustness of the helical edge channels for different temperatures. Figures 4A-C represent the local resistances $R_{03,12}$ for Hall bar devices with $L_L$ = 3, 2, and 1 µm at increasing temperatures. The gate-training technique has been employed at each temperature. $R_{max}$ is presented as a function of temperature in Fig. 4D. $R_{max}$ remained rather constant at $R_{max}$ = h/2e² until they reached their respective threshold temperatures $T_c$. These are: $T_{c,3µm}$ = 40 K, $T_{c,2µm}$ = 60 K and $T_{c,1µm}$ = 60 K, while it seems to be the most robust for $L_L$ = 1 µm. This robustness of the helical edge channels for smaller devices is further supported by a model of edge resistance with additional backscattering in Fig. S12 in the Supplementary Materials. The onset temperature of quantized conductance was extracted as in Ref. (*17*) with a maximal deviation of up to 15%. The observation of conductance quantization up to 40 K can be attributed to the increase in band gap energy, and subsequent reduction of bulk conductivity. This maximum operation temperature could be reproducibly achieved in the devices investigated. However, the achieved quantization up to 60 K, as presented for the 1 µm and 2 µm devices, stems from device-to-device variations in the residual bulk conductivity and subsequent larger bulk resistance as obtained from the large-scale Hall bar data (see Fig. S6 in the Supplementary Materials). These variations could be due to growth or processing, which requires further investigations. With improvements in sample growth and fabrication, these variations can be improved in the future. The microscopic devices further exhibit a low mobility (see. Fig. S13 in the Supplementary Materials) leading to an overall larger bulk resistivity. We also emphasize that since the resistance at the maximum operation temperature remains close to the quantized value for all devices, the bulk resistance must still exceed the helical edge resistance, indicating that transport is dominated by the edges.

Additionally, the overall maximum temperature, at which the QSHE in the device with $L_L$ = 1 µm is observed, is not related to the deviation from the expected resistance caused by the increased backscattering in the helical edge channels. Instead, the plateau shifts to lower gate voltages until the band gap region cannot be probed in the up-sweep for the optimized $V_{min}$ anymore (grey region), which is important to optimally apply the gate training. This shift of the whole curve can be attributed to pronounced hysteresis effects arising between the up- and down-sweep during the gate-training process, which limits the applicability of the gate training at elevated temperatures (see Fig. S15 in the Supplementary Materials). The hysteresis originates from charge accumulation at the interface between the gate dielectric and the cap layer of QW heterostructures (*40, 41*). These hysteresis effects are also the reason why higher temperatures for the 3 µm device are not shown compared to the other devices as the gap region in the up-sweep could not be probed at $T = 50$ K and 60 K anymore. We anticipate that with improvements in device quality, especially regarding the stability of the gate, it will be possible to overcome these hysteresis issues. Exploring other capping layers or high-k dielectrics such as $Al_2O_3$ or $HfO_2$ might also mitigate this issue. In addition to improving the device quality, it is necessary to fabricate QW heterostructures with higher inverted band gap values for the reliable observation of the QSHE at even higher temperatures. In particular, the band gap of the $InAs/Ga_{0.68}In_{0.32}Sb/InAs$ TQW used in this work has a moderate value of 27 meV. By growing $InAs/Ga_{0.68}In_{0.32}Sb/InAs$ TQWs with slightly different layer thicknesses, the band gap can be increased up to ~42 meV (see Fig. 1). Moreover, by utilizing the $Ga_{0.60}In_{0.40}Sb$ alloy and adjusting the buffer materials, which are feasible parameters for high-quality pseudomorphic growth (*51*), the band gap in the TQW exceeds 50 meV at realistic strain conditions in the layers (see Fig. S14 in the Supplementary Materials). Thus, our experimental results obtained in InAs/GaInSb/InAs TQWs with a moderate band gap make these QW structures extremely attractive for advancing topological electronics above liquid nitrogen temperatures.

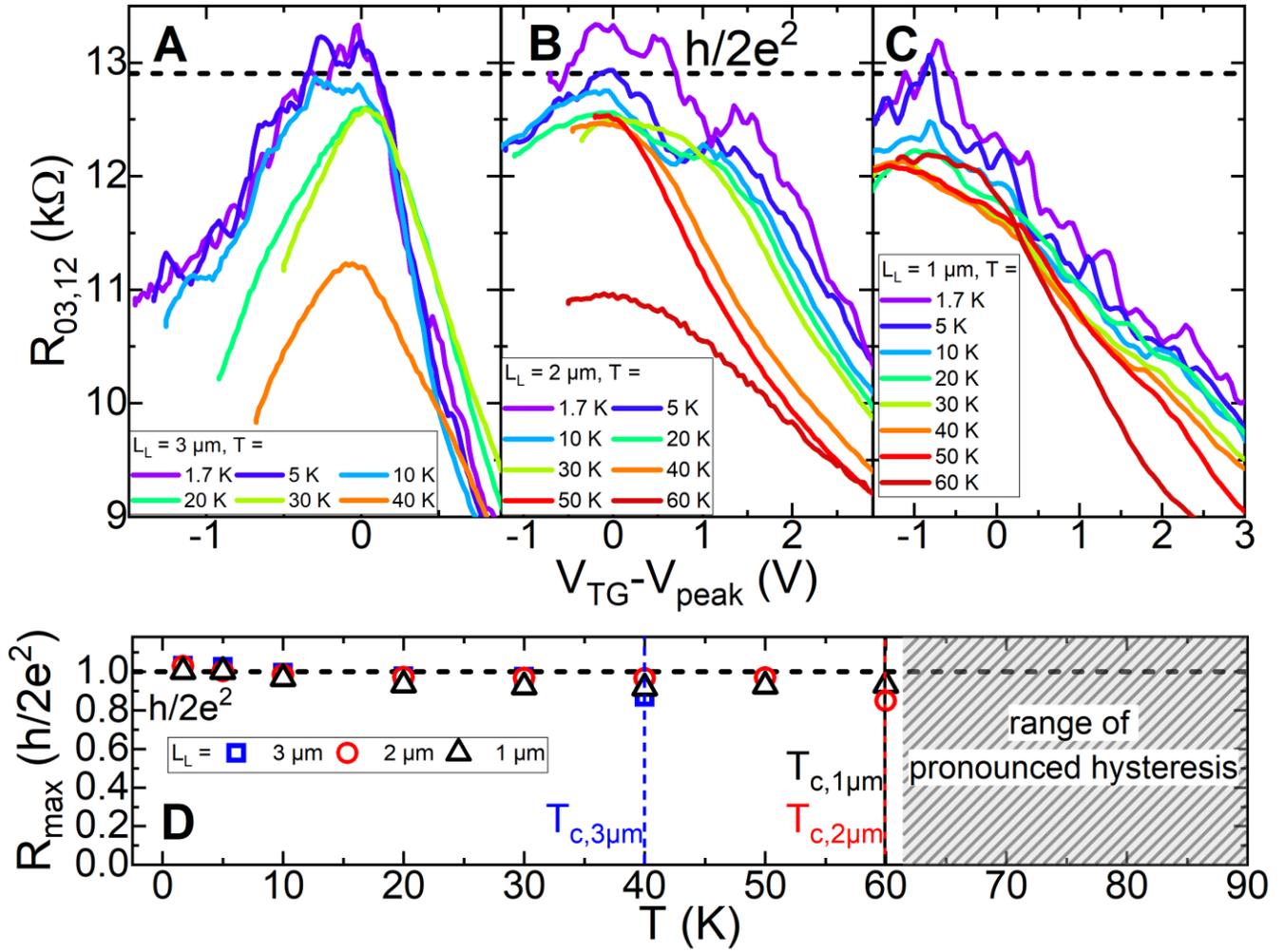

**Fig. 4. Robust conductance quantization up to T = 60 K.** Temperature dependence of the quantized resistance value in the band gap for $L_L$ = 3, 2 and 1 µm in **(A)**, **(B)**, and **(C)**, respectively. The resistance maximum values are summarized in panel **(D)**. For all three devices, a constant $R_{max}$ at $h/2e^2$ can be observed with a maximum temperature of $T$ = 60 K for the 1 µm and 2 µm Hall bar devices. At larger temperatures, the gate-training range exceeds the accessible voltage range, at which the gate operates stably (greyed region).

**DISCUSSION**

In conclusion, we demonstrated the QSHE up to $T$ = 60 K in a material system suitable for device applications. The length-independent quantized resistance values for the gap observed in both local and nonlocal measurement geometries are striking evidence of the QSHE. Utilizing the gate-training technique in devices with a deviating resistance from the expected value, a quantized resistance value could be approached. Furthermore, our temperature study revealed that electronic transport is clearly dominated by helical edge channels with stable quantized resistance up to $T = 60$ K, while observations at even higher temperatures have been prohibited due to pronounced hysteresis effects and the moderate band gap energy. Considering the continuous improvements in material and device quality and by increasing the band gap energy, we conclude that achieving the QSHE in this material system at temperatures above that of liquid nitrogen is becoming realistic. Our findings pave the way for developing topological electronics technology at non-cryogenic temperatures utilizing III-V semiconductors with well-established growth and processing technologies and enabling integration with silicon technology.

# MATERIALS AND METHODS

## Sample growth and fabrication

The sample was grown by molecular beam epitaxy (MBE) on an n-doped (001) GaSb substrate, followed by an undoped 200 nm GaSb buffer. Subsequently, a 1500 nm AlSb quasi-substrate was grown to change the lattice constant from GaSb to AlSb. This is followed by a 10x (2.5/2.5) nm GaSb/AlSb superlattice to reduce the dislocation density. The TQW consists of two 9 nm InAs layers separated by a 3.1 nm $Ga_{0.68}In_{0.32}Sb$ layer and sandwiched between two 40 nm AlSb barriers. A 5 nm GaSb cap was grown on top of the sample to protect it against oxidation. All microscopic Hall bar devices in this study have been fabricated from the same QW heterostructure. For all lithography steps, e-beam lithography was used. As for the etching of the Hall bars, conventional dry-etching techniques were used via reactive ion etching with Ar and Cl. For the ohmic contacts, all antimonide-containing layers are selectively etched using the wet-chemical etchant tetramethylammonium hydroxide (TMAH). The InAs layer is then contacted with Ohmic contacts consisting of Cr and Au. As for the gate dielectric, a superlattice of $SiO_2$/SiN (5x 10 nm/10nm and ending with an additional 10nm $SiO_2$ layer) is applied using plasma-enhanced chemical vapor deposition. More information can be found in Fig. S1 in the Supplementary Materials.

## Band structure calculations

Band structure calculations have been performed by using the eight-band k·p Hamiltonian (*24*), which directly considers the interactions between $\Gamma_6$, $\Gamma_8$ and $\Gamma_7$ bands in bulk materials. In the Hamiltonian, we also consider the terms describing the strain effects arising due to mismatch of lattice constants in the buffer, QW layers and AlSb barriers. The calculations have been performed by expanding the eight-component envelope wave functions in the basis set of plane waves and by numerical solution of the eigenvalue problem. More details of calculations and the form of the Hamiltonian can be found in Ref. (*24*) and the Supplementary Materials. Parameters for the bulk materials and valence band offsets used in the calculations are taken from Ref. (*52*).

**Acknowledgments**

**Funding:** The work was supported by the Elite Network of Bavaria within the graduate program "Topological Insulators" and by the Occitanie region through the TOP platform and the "Quantum Technologies Key Challenge" program (TARFEP project). We acknowledge financial support from the DFG within the project HO 5194/19-1 and through the Würzburg-Dresden Cluster of Excellence on Complexity and Topology in Quantum Matter – ct.qmat (EXC 2147, project-id 390858490). We also acknowledge the French Agence Nationale pour la Recherche (Cantor project - ANR-23-CE24-0022), and Equipex+ Hybat project - ANR-21-ESRE-0026), and the Physics Institutes of CNRS for Emergence 2024 - Step - project. We are grateful for the support of this work by the State of Bavaria.

# Supplementary Materials for
## 'Quantum spin Hall effect in III-V semiconductors at elevated temperatures: advancing topological electronics'

**Layer structure and sample fabrication**

The layer structure of the sample investigated is provided in Fig. S1A. The sample was grown by molecular beam epitaxy (MBE) on an n-doped (001) GaSb substrate, followed by an undoped 200 nm GaSb buffer.

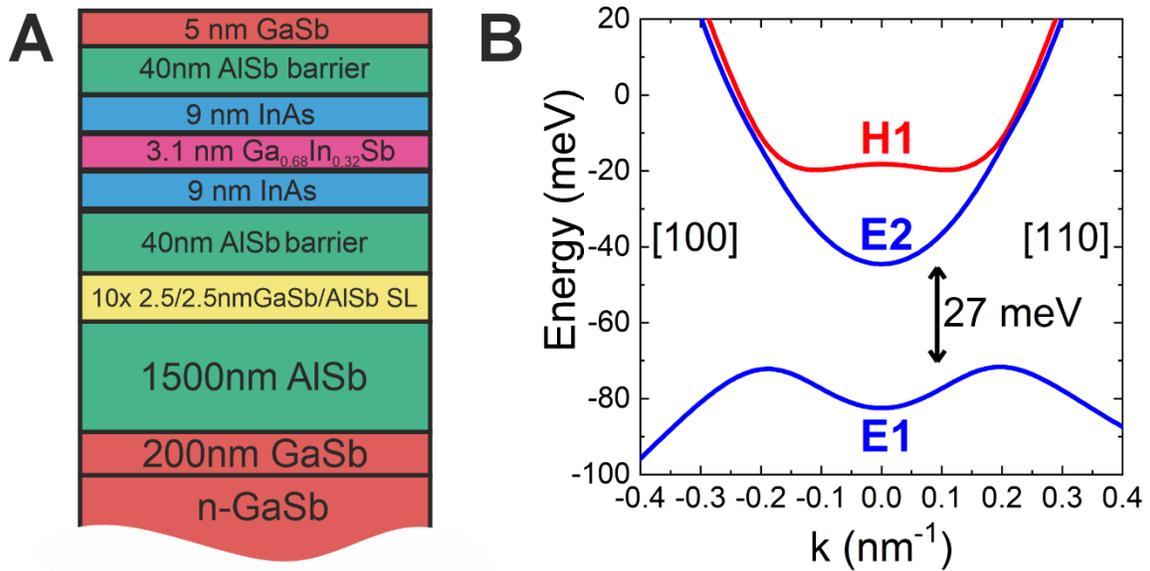

**Fig. S1.** (**A**) Layer structure of the symmetric InAs/GaInSb/InAs trilayer quantum well (TQW) grown on (001) AlSb buffer. (**B**) Band structure of the grown TQW. The blue and red curves represent the energy-dispersion of the electron-like (E1, E2) and hole-like (H1) subbands, respectively. The positive and negative values of quasimomentum $k$ correspond to the [100] and [110] crystallographic orientations. The indirect band gap in TQW is about 27 meV.

Subsequently, a 1500 nm AlSb quasi-substrate was grown to change the lattice constant from GaSb to AlSb. This is followed by a 10x (2.5/2.5) nm GaSb/AlSb superlattice to reduce the dislocation density. The TQW consists of two 9 nm InAs layers separated by a 3.1 nm $Ga_{0.68}In_{0.32}Sb$ layer and sandwiched between two 40 nm AlSb barriers. A 5 nm GaSb cap was grown on top of the sample to protect it against oxidation. All microscopic Hall bar devices in this study have been fabricated from the same QW heterostructure. For all lithography steps, e-beam lithography was used. As for the etching of the Hall bars, conventional dry-etching techniques were used via reactive ion etching with Ar and Cl. For the ohmic contacts, all antimonide-containing layers are selectively etched using the wet-chemical etchant tetramethylammonium hydroxide (TMAH). The InAs layer is then contacted with Ohmic contacts consisting of Cr and Au. As for the gate dielectric, a superlattice of $SiO_2$/SiN (5x 10 nm/10nm and ending with an additional 10nm $SiO_2$ layer) is applied using plasma-enhanced chemical vapor deposition.

Fig. S1B represents the energy band dispersion of the grown symmetrical TQW with an indirect bandgap between the E2 and E1 bands of $E_{gap} = 27$ meV.

## Experimentally extracted band gap value

To confirm the theoretically predicted bandgap value, macroscopic Hall bars were fabricated, and temperature-dependent measurements were performed. The contact separation length of the Hall bar is $L_L = 10$ μm and the complete length and width are $L = 130$ μm and $W = 20$ μm. In Fig. S2A, the longitudinal resistance $R_{03,12}$ versus $T = 1.7$ K to 130 K is presented. From the Arrhenius plot in Fig. S2B, the band gap energy can be extracted to $E_{gap} = (24.2 \pm 2.8)$ meV by applying a linear fit to the high-temperature regime (19). This experimental gap value is in good agreement with the one extracted from the calculations of 27 meV.

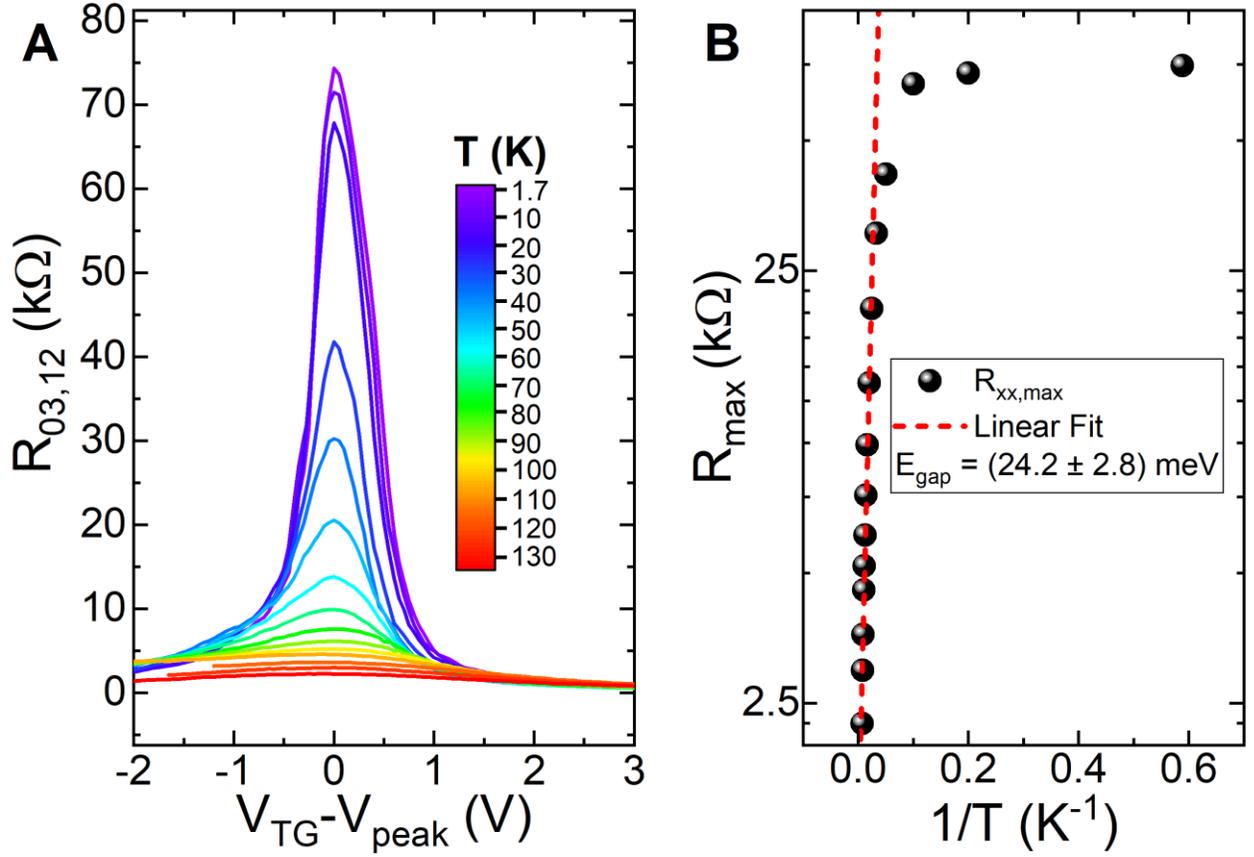

**Fig. S2.** (**A**) Temperature dependence of the resistance $R_{03,12}$ for $T$ = 1.7 to 130 K for a macroscopic Hall bar. (**B**) Arrhenius plot of the resistance maxima. By applying a linear fit to the high-temperature regime, the band gap can be extracted to $E_{gap}$ = (24.2 ± 2.8) meV.

### Expected quantized values in local and nonlocal configurations

In the following, we show that the four-probe resistances $R_{03,12}$ and $R_{01,32}$ are given by $R_{03,12} = h/2e^2$ and $R_{01,32} = h/6e^2$ for helical edge states without backscattering, using the Landauer-Büttiker formalism.

We use the well-known formula for multiterminal conductors (*53*):

$$I_p = \sum_q G_{pq}(V_p - V_q),$$

Where $I_p$ is the current flowing from the external circuit into the lead $p$, $V_q$ is the voltage at the lead $q$ and $G_{pq}$ is the conductance from lead $p$ to lead $q$. In the simplest case of perfect ballistic helical edge states, for the geometry indicated in Figure S8, the conductances $G_{pq}$ are given by the elements of the conductance matrix:

$$G = \begin{bmatrix} 0 & 1 & 0 & 0 & 0 & 1 \\ 1 & 0 & 1 & 0 & 0 & 0 \\ 0 & 1 & 0 & 1 & 0 & 0 \\ 0 & 0 & 1 & 0 & 1 & 0 \\ 0 & 0 & 0 & 1 & 0 & 1 \\ 1 & 0 & 0 & 0 & 1 & 0 \end{bmatrix}, \quad (1)$$

where the unit of conductance has been taken equal to 1. Let us calculate the resistances for a current $I_{03}$ flowing from lead 0 to lead 3. Setting $V_3 = 0$ and leaving out the row and column corresponding to lead 4, we get the linear system:

$$\begin{bmatrix} I_{03} \\ 0 \\ 0 \\ 0 \\ 0 \end{bmatrix} = \begin{bmatrix} 2 & -1 & 0 & 0 & -1 \\ -1 & 2 & -1 & 0 & 0 \\ 0 & -1 & 2 & -1 & 0 \\ 0 & 0 & -1 & 2 & -1 \\ -1 & 0 & 0 & -1 & 2 \end{bmatrix} \begin{bmatrix} V_0 \\ V_1 \\ V_2 \\ V_4 \\ V_5 \end{bmatrix}, \quad (2)$$

From which all the voltages of the different probes are easily calculated. From equation (2), the resistance $R_{03,12}$ can be calculated to $R_{03,12} = (V_2 - V_1) / I_{03} = 1/2$ (Fig. S3A). A similar calculation can be done for $R_{01,32}$. It yields $R_{01,32} = (V_3 - V_2) / I_{01} = 1/6$ (Fig. S3B).

Such calculations can be performed for much more complex devices, introducing 'virtual contacts' between the real ones, or assuming some backscattering between the 'real' leads.

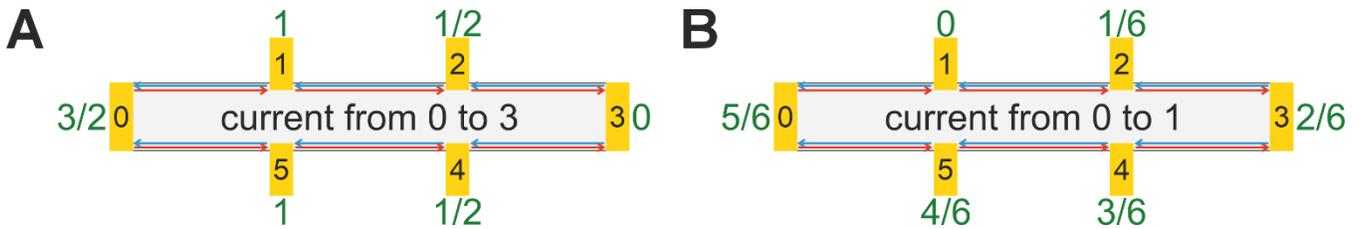

**Fig. S3.** Scheme of a Hall bar with contacts labeled from 0 to 5 and the helical edge states are shown as red and blue arrows. The voltages are indicated in green, in units of $h/e^2$. (**A**) The current flows from lead 0 to lead 3. (**B**) The current flows from lead 0 to lead 1.

## Reproducibility of the peak positions

To confirm the reproducibility of the peak position, experiments in local configuration $R_{03,12}$ with $L_L = 3$ µm and nonlocal configuration $R_{01,32}$ with $L_{NL} = 3.5$ µm have been performed. These are depicted in Figs. S4A and B, respectively. Three measurements for each configuration were conducted for the same gate-voltage range ($V_{TG} = +3$ to $-3$ V). The resistance peaks always remain at the same position of $V_{TG} \approx -1.47$ V.

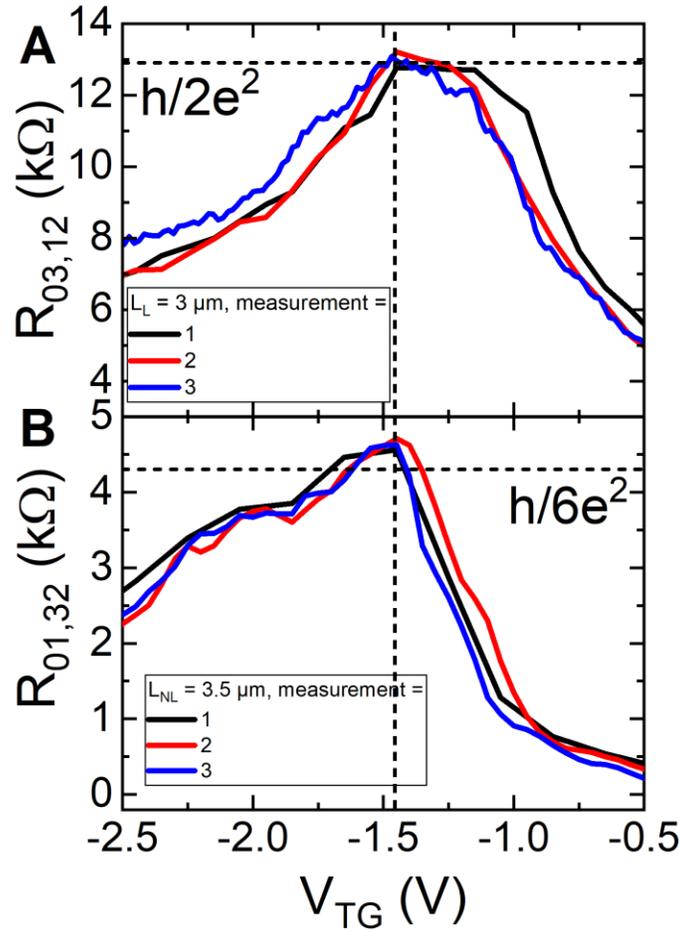

**Fig. S4.** Peak position of the same device for the local configuration $R_{03,12}$ with $L_L = 3$ µm in (**A**) and for the nonlocal configuration $R_{01,32}$ with $L_{NL} = 3.5$ µm in (**B**). For both configurations, the measurements have been performed three times, and the peak can always be found at $V_{TG} \approx -1.47$ V.

# Additional measurement configurations

As shown before, dependent on the measurement configuration (i.e., where the current is applied and the voltage is measured), different quantized resistance values are expected. In Fig. S5A, the current is applied between contacts 1 and 5 and the resistance is measured between 2 and 3. The resistance value for the gap is in good agreement with the expected value of $h/3e^2$. In Fig. S5B, the current is applied between contacts 0 and 5 and the voltage is either measured between 2 and 3 or 3 and 4. In both cases, the expected quantized value of $h/6e^2$ again coincides with the maximum resistance in the gap region. These additional configurations are further evidence for the QSHE.

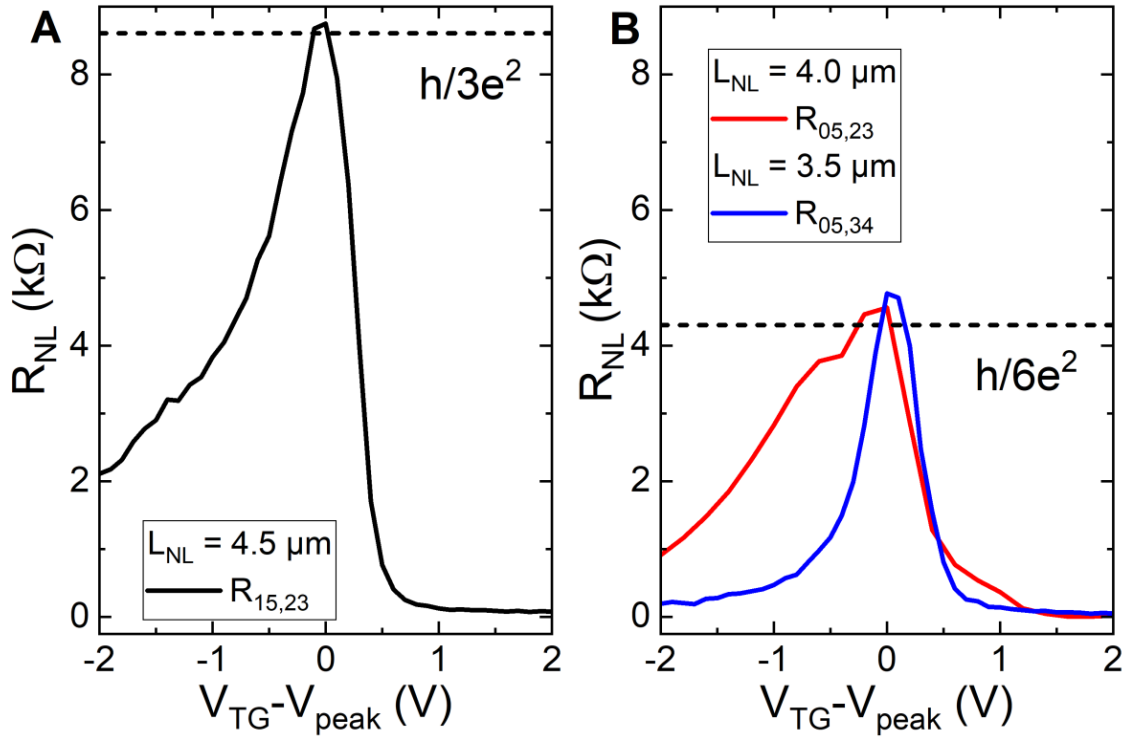

**Fig. S5.** (**A**) $R_{15,23}$ as a function of $V_{TG}$ for a device with $L_{NL}$ = 4.5 µm. The maximum resistance in the gap is in good agreement with the expected value of $h/3e^2$. (**B**) $R_{05,23}$ and $R_{05,34}$ as a function of $V_{TG}$ for a device with $L_{NL}$ = 4.0 µm and 3.5 µm, respectively. The maximum resistances are in both cases in good agreement with the expected value of $h/6e^2$.

**Separating bulk and edge contributions**

To observe the QSHE, trivial parasitic contributions need to be negligible. In InAs/GaSb BQWs, the low band gap (*18, 30*) and native p-doping (*54*) of antimonides resulted in a large residual bulk contribution and impeded the observation of the QSHE. The bulk resistance of BQWs or TQWs could be improved by doping the interface between InAs and GaSb with Si (*19*) or by growing GaInSb instead of GaSb (*36, 38, 39*). The question remains how much this improves the bulk insulation and if the current flows solely at the edges due to the QSHE. Therefore, we performed a multi-probe analysis on macroscopic devices to separate bulk and edge contributions. Detailed information about this method can be found in Ref. (*46*). Briefly, this technique consists in the measurement of 45 four-probe resistances as a function of top-gate voltage. Then, for each gate voltage of interest, the 45 resistances are fitted with only two parameters, the bulk conductivity (in µS/sq, sq = 1 µm × 1 µm) and the edge conductance (in µS/µm). The edge conductivity $G_e$ and bulk conductivity $\sigma_B$ for two different devices HB10a and HB70a are shown in Fig. S6A and B, respectively. The dimensionalities of the Hall bars are $W = 20$ µm and $L_L = 10$ and $70$ µm for HB10a and HB70a, respectively. From the analysis of the HBs in the band gap, one observes that the bulk conductivity becomes negligibly small, indicating an insulating bulk with a resistivity of approx. 2 MΩ/sq for HB10a and even more than 100 MΩ/sq for HB70a.

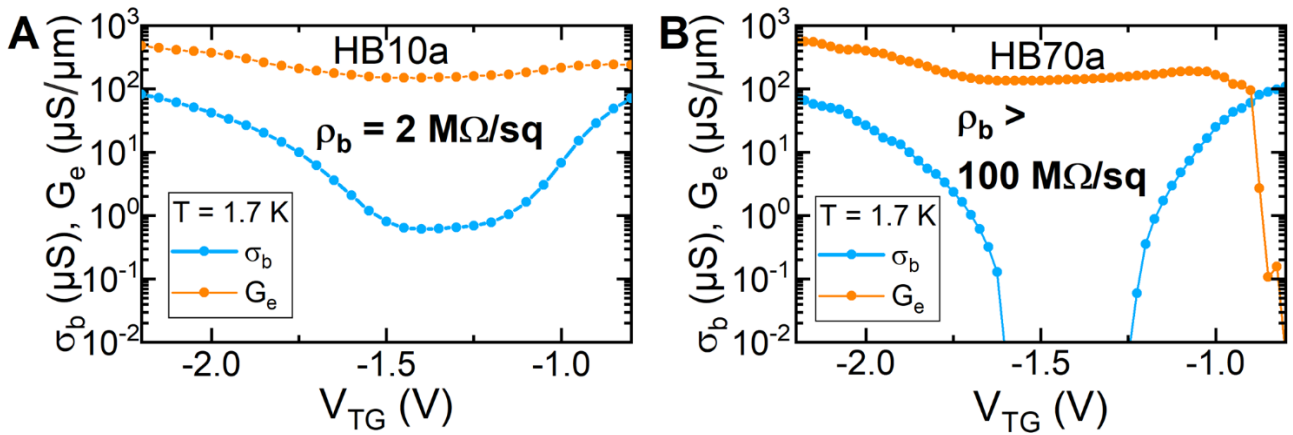

**Fig. S6.** Evolution of the edge conductance $G_e$ (orange dots) and bulk conductance $\sigma_b$ (blue dots) as a function of top-gate voltage $V_{TG}$ for HB10a ($L_L$ = 10μm) in (**A**) and HB70a ($L_L$ = 70 μm) in (**B**), respectively. In the gap, (around $V_{TG}$ = -1.2V) the bulk conductance is diminished, and the transport is dominated by the edges.

The difference of these bulk resistivity values can be explained by a variation of the number of defects. In general, the residual bulk conductivity in non-intentionally doped GaSb-based materials grown by MBE results mostly from native Ga antisite defects and/or Ga vacancy p-doping defects (*54–56*), whereas the AlSb quasi substrate is insulating (*57*). Additional conductivity paths may arise from dislocations due to the growth on the AlSb buffer, as dislocations are known to generate carrier leakage in III-V semiconductor heterostructures (*58–60*). For both devices, the bulk conductance is diminished, and the transport is dominated by the edges. With $G_e$ = 150 μS/μm for both devices, one can extract the phase coherence length to $\lambda = (G_e h/e^2) \approx 3.9$ μm (*46*), which is in good agreement with the value extracted from Fig. 2C from the manuscript.

## Determination of the phase coherence length

To accurately extract the phase coherence length $\lambda$, we performed a multi-probe analysis (*46*) on a few Hall bar devices to extract the edge resistance $R_{edge}$ as a function of contact separation length. In Fig. S7A we show the (local) longitudinal resistances $R_{03,12}$ measured in HB10a at $T$ = 5 K for the down sweep (in blue) and up sweep (in green). The resistance depends on the direction of the gate voltage sweep. From Fig. S6, we know that the bulk conductivity is negligible at these low temperatures. Hence, the maxima of the $R_{03,12}$ peaks already give a first estimate of the backscattering length. Using $R_{03,12} \simeq \frac{h}{2e^2} L/\lambda$, a first estimation yield $\lambda \simeq 3 - 5\ \mu m$. For more precise determination of $\lambda$, we performed four-probe resistance measurements in all 45 geometries as a function of top gate voltage. Then the 45 maxima of the resistance peaks were fitted with six resistances $R_{edge}$ of the six edges, as independent parameters. The edge lengths are 10 μm (two edges) and 50 μm (four edges) for HB10a. The fits are very good, even if one considers the value of the reduced chi-square – it means that the model is meaningful. The six obtained edge resistances are reported in Fig. S7B for the two directions of the gate voltage sweeps. From these resistances, the backscattering length can be fitted as $R_{edge} = \frac{h}{e^2}\frac{L}{\lambda}$, which gives $\lambda \simeq 4 - 6$ $\mu m$. All these values suggest that a quantization should be observed in micrometer-sized devices. Note that since the gate training method was not applied during this experiment, one can expect a larger $\lambda$ after gate training as it is already indicated by the difference between the two sweep directions.

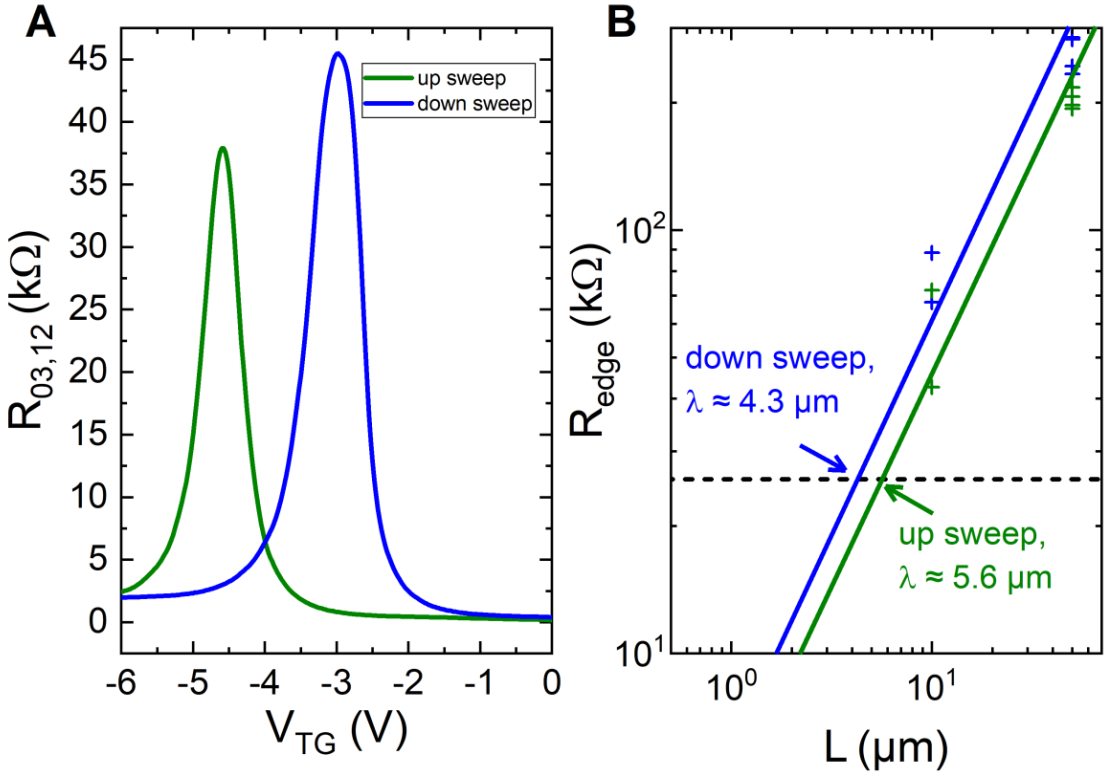

**Fig. S7**. Macroscopic device HB10a measured at $T = 5$ K. (**A**) Longitudinal resistance $R_{03,12}$ for the down sweep (in blue) and up sweep (in green), respectively. (**B**) The symbols are the results of the 6-parameter fits for the edge resistance in dependence of the edge lengths.

The same analysis was applied to other devices (4 in total). We therefore obtain the edge resistances for different lengths of the different devices, as shown in Fig. S8. The green triangles are the extracted edge resistances from the data, and the solid line is the best fit with $R_{edge} = \frac{h}{e^2}\frac{L}{\lambda}$. We emphasize that $\lambda$ in Figs.

S7B and S8 is obtained at the intersection of the fit with $h/e^2$ and not $h/2e^2$ as the fit gives the edge resistances separately. We get $\lambda \approx 4.7$ μm, a value greater than all the edge lengths of the microscopic devices.

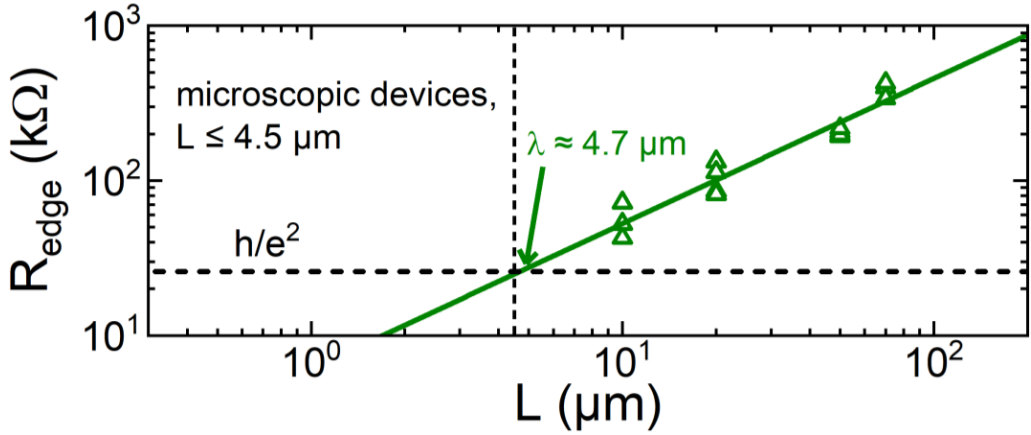

**Fig. S8.** $R_{edge}$ extracted from a six-parameter fit as a function of edge length for several Hall bar devices. The intersection with $h/e^2$ gives the phase coherence length $\lambda \approx 4.7$ μm.

Remarkably, the gate training technique was also not even used in these experiments. One might therefore expect an even larger $\lambda$ after gate training.

**Breaking of time-reversal symmetry and helical edge states**

The robustness of the QSHE in the presence of magnetic field strongly depends on whether a 2D system possesses spatial inversion symmetry, as well as on the orientation of the applied magnetic field. Indeed, the presence of an inversion center leads to the conservation of physical spin $S_z$, where $z$ is the axis perpendicular to the plane of the 2D system. In this case, the polarization of the edge states is also oriented along the $z$ axis; and the application of a perpendicular magnetic field does not lead to the breaking of the QSHE until the field exceeds a critical value $B_c$, in which the zero-mode Landau level cross (*61*). On the contrary, the application of even small magnetic fields in the 2D system plane leads to the gap opening at the Dirac point of the edge states. If the Dirac point of the edge states is located inside the band gap of the bulk states, then this leads to the breaking of the QSHE. However, if the Dirac point is buried inside conduction or valence band of the bulk states, the quantum edge transport remains robust even in the presence of in-plane magnetic field (*62*). Importantly, the presence of inversion asymmetry yields the gap opening for the edge states already in a perpendicular magnetic field. While an applied vertical electric field through the top gate breaks structure inversion asymmetry (SIA) also in the TQWs, in the gap region the electric field is rather small. Figure S9 provides measurements of the longitudinal resistance $R_{03,12}$ under small magnetic field applied in-plane (Fig. S9A) and out-of-plane (Fig. S9B) of the TQW at low temperatures.

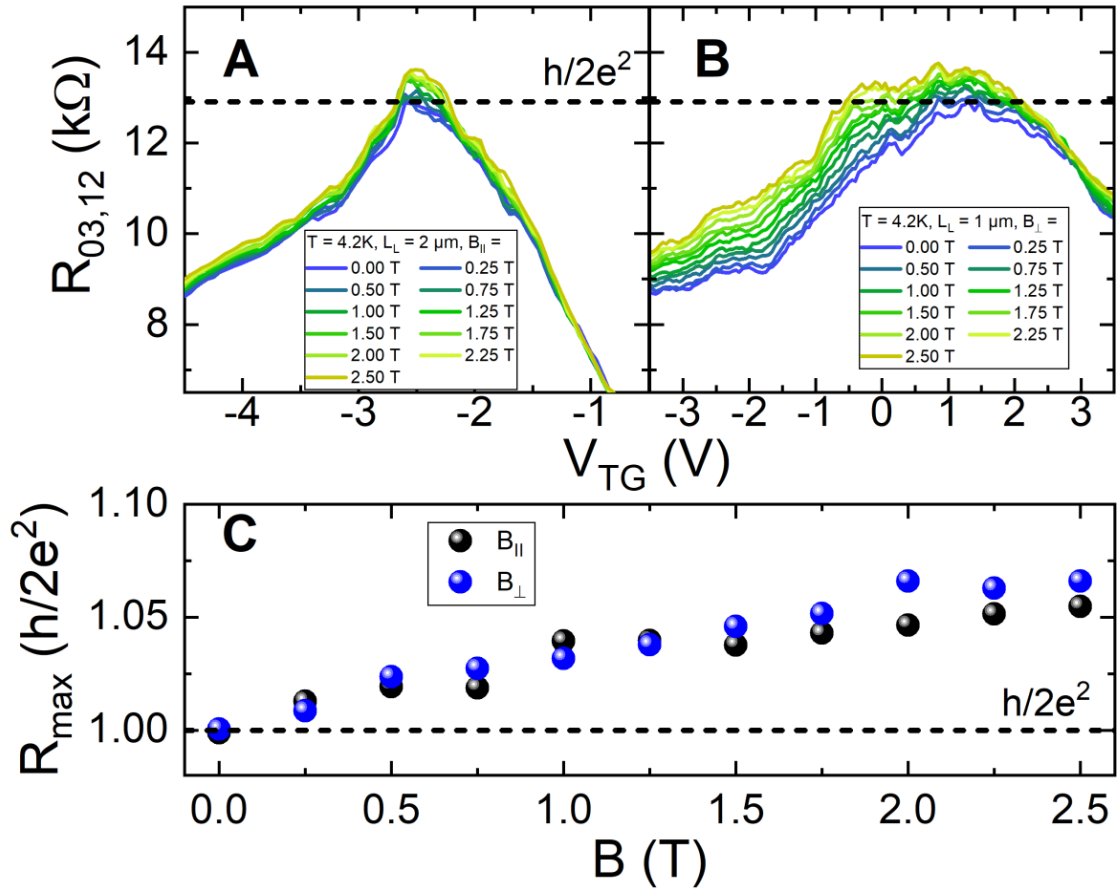

**Fig. S9.** Longitudinal resistance $R_{03,12}$ as a function of top-gate voltage $V_{TG}$ at $T = 4.2$ K for (**A**) in-plane magnetic field $B_{\parallel}$ (parallel to the current) and (**B**) perpendicular magnetic field $B_{\perp}$. (**C**) The resistance maxima as a function of magnetic field for both field configurations.

As seen in Fig. S9C, the maximum resistance in the band-gap region increases with magnetic field that indicates the gap opening for the edge states, suppressing of the quantized values of the QSHE for both magnetic field configurations. The latter evidences a pronounced role of the bulk inversion asymmetry (BIA) and the interface inversion asymmetry (IIA) in the symmetric InAs/GaInSb/InAs TQW. We note that a similar behavior of the longitudinal resistance, but with a more pronounced effect of change at the same values of the magnetic field was also observed in an InAs/GaInSb BQWs (*36*). The stronger change in $R_{03,12}$ may indicate the larger g-factor values for the edge electrons due to the presence of SIA inherent in InAs/GaInSb BQWs. Alternatively, considering the band structure of the grown sample, the weakness of the magnetic field effect on $R_{03,12}$ may be because the Dirac point of the edge states is hidden inside either the conduction E2 subband or the valence E1 subband (see Fig. S10A). In this case, the effect of the magnetic field on longitudinal resistance is expected to be substantially stronger for the sample with the maximum band gap, for which the E2 and H1 subbands are located close in energy (see Fig. S10B).

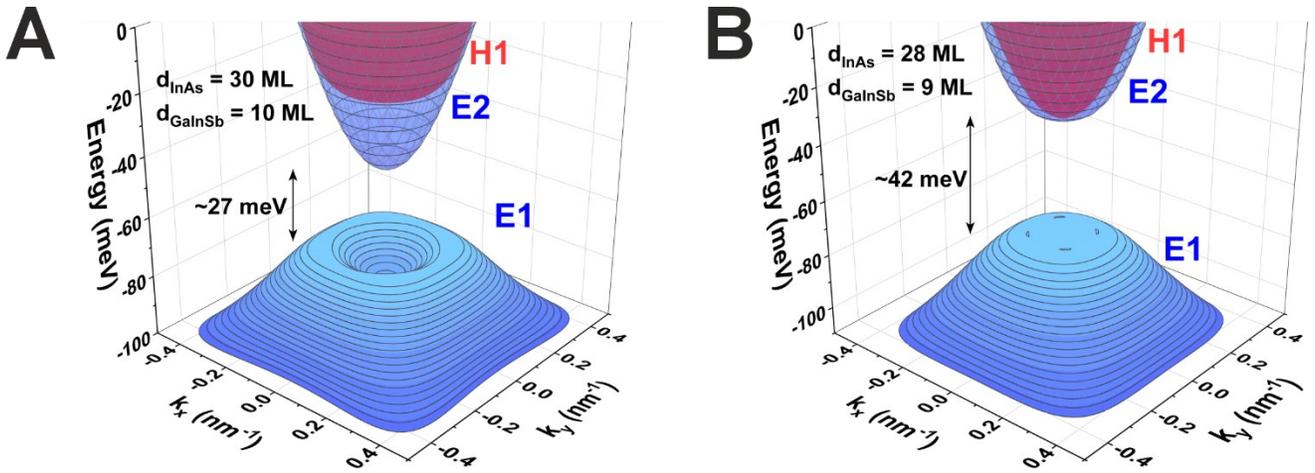

**Fig. S10.** 3D plots of the band structure for the sample from the main text in (**A**) and the sample with the highest possible band gap for Ga$_{0.68}$In$_{0.32}$Sb on AlSb in (**B**). The *x* and *y* axes are oriented along [100] and [010] crystallographic directions, respectively.

Indeed, calculations (see Fig. S11) show that the Dirac point of the helical edge states is likely buried within the conduction E2 subband rather than lying within the bulk gap. In real devices, although boundary conditions may vary, it is reasonable to assume that the Dirac point remains hidden inside the bulk band, which reduces the magnetic field's ability to open a sizable gap at the Dirac point and thus suppressing edge conduction.

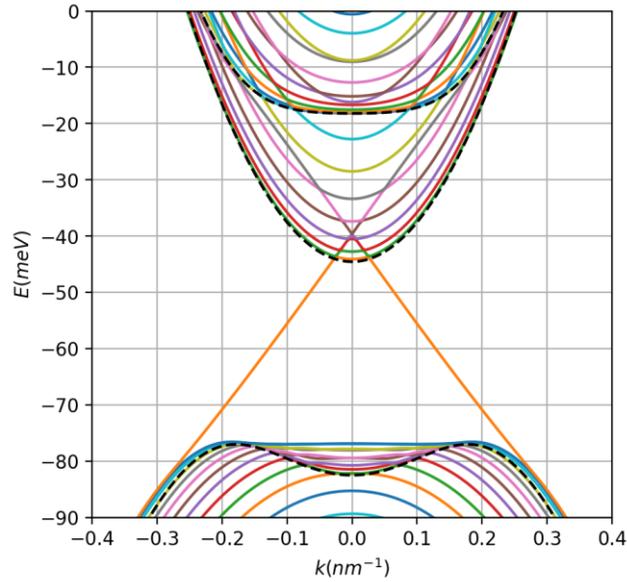

**Fig. S11.** Band dispersion for a nanoribbon made from the same trilayer structure (solid lines: nanoribbon; dashed black lines: bulk dispersion). The Dirac point is buried in the conduction band. Calculation performed with the kwant package (tight binding calculations).

To clarify the reasons for such a weak effect of the magnetic field on $R_{03,12}$ in our devices compared to the previously studied BQWs is beyond the scope of this paper and will be the subject of future investigations. Nevertheless, the small suppression of quantized $R_{03,12}$ values by the magnetic field in microscopic Hall bar devices shown in Fig. S9 can be still considered as a breaking of TRS by magnetic field.

## Model of the edge resistance with additional backscattering

We assume now that the edge resistance between two leads separated by a length $L$ is given by the simple formula:

$$R_e(L) = \max\left(1, \frac{L}{\lambda}\right), \tag{3}$$

where $\lambda$ is the phase coherence length. This formula mimics the saturation of the resistance in the shortest edges, as well as their ohmic behavior in the longest edges. The conductance matrix is then given by:

$$G = \begin{bmatrix} 0 & R_e(L_{NL}) & 0 & 0 & 0 & R_e(L_{NL}) \\ R_e(L_{NL}) & 0 & R_e(L_L) & 0 & 0 & 0 \\ 0 & R_e(L_L) & 0 & R_e(L_{NL}) & 0 & 0 \\ 0 & 0 & R_e(L_{NL}) & 0 & R_e(L_{NL}) & 0 \\ 0 & 0 & 0 & R_e(L_{NL}) & 0 & R_e(L_L) \\ R_e(L_{NL}) & 0 & 0 & 0 & R_e(L_L) & 0 \end{bmatrix}, \tag{4}$$

The results of the calculation of the local and non-local resistances for different geometries are given in Figure S12. At $\frac{1}{\lambda} \simeq 0$, all resistances are quantized. Increasing $\frac{1}{\lambda}$, the nonlocal resistance $R_{01,32}$ loses its quantization 1/6 first for devices $L_L$=1 (the largest $L_{NL}$), then for device $L_L$=2 and finally for device $L_L$= 3 μm. The resistance quantization is more robust in the local configuration $R_{03,12}$, but is also lost when $L_L$ exceeds $\lambda$. It is the most robust for $L_L = 1$ μm.

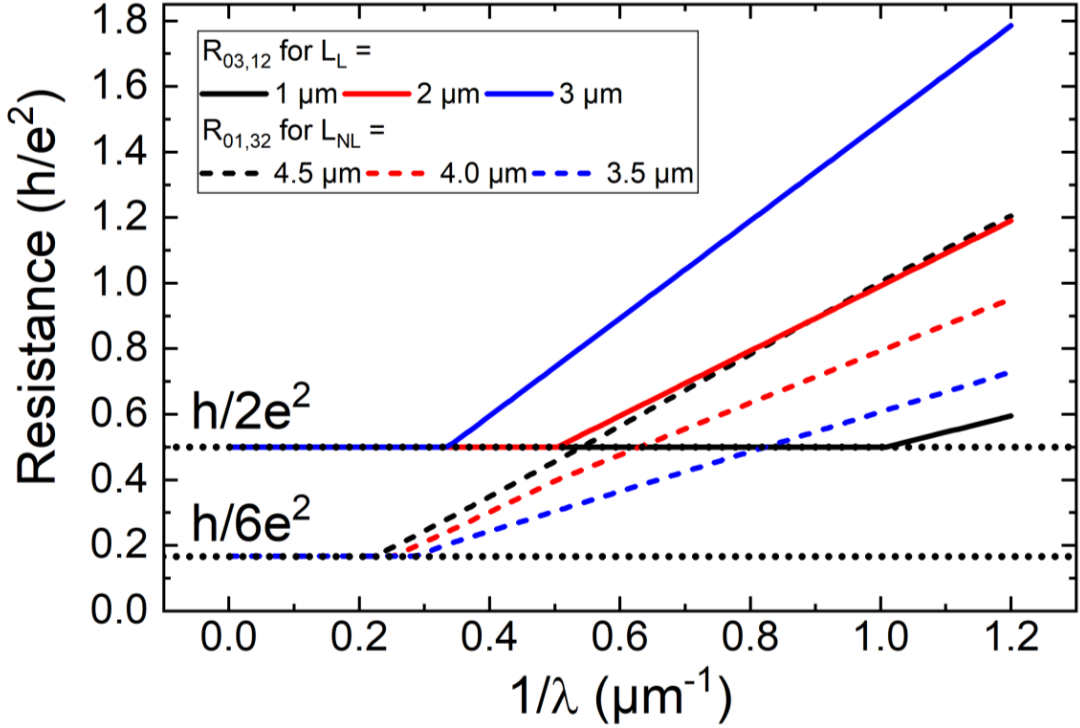

**Fig. S12.** Resistance as a function of $1/\lambda$ in local and nonlocal configurations. For the nonlocal configuration ($R_{01,32}$), the quantized value of the resistance is first lost for $L_L = 1\,\mu m$ ($L_{NL} = 4.5\,\mu m$), then for $L_L = 2\,\mu m$ ($L_{NL} = 4.0\,\mu m$) and at last for $L_L = 3\,\mu m$ ($L_{NL} = 3.5\,\mu m$). In the local configuration ($R_{03,12}$), the quantized value of the resistance is more robust and is at last lost for $L_L = 1\,\mu m$.

## Mobility and mean free path

One can also expect a quantized resistance for 1D edge channels if the contact separation length L between the voltage probes is smaller than the mean free path $\lambda_{mfp}$. This would require 1D edge channels in our 2D system even without the quantum spin Hall effect (QSHE), which is unlikely. However, we still want to exclude this possibility of the quantization for $L < \lambda$ for our devices studied in the main text. Figure S13 shows an additional Hall measurement for the Hall bar device with $L_L = 2\,\mu m$ from Figure 2 in the main text. From this measurement, the charge carrier density $n$, mobility $\mu$ and $\lambda_{mfp} = \frac{\hbar}{e}\mu\sqrt{2\pi n}$ (20) was extracted at a given top-gate voltage $V_{TG} = +10$ V, which yield $n = 9.31 \times 10^{11}$ cm$^{-2}$, $\mu = 13.4 \times 10^3$ cm$^2$/Vs and $\lambda_{mfp} = 107$ nm. These values are extracted when the Fermi energy lies deep in the conduction band and even there $\lambda_{mfp}$ is smaller than all the contact separation lengths. In the band gap, where the helical edge channels appear, the mean free path is expected to be much smaller because of smaller charge carrier densities and mobilities. Therefore, the observed quantized resistances in the main text are attributed to the observation of the QSHE. The overall low mobility also indicates a larger bulk resistivity.

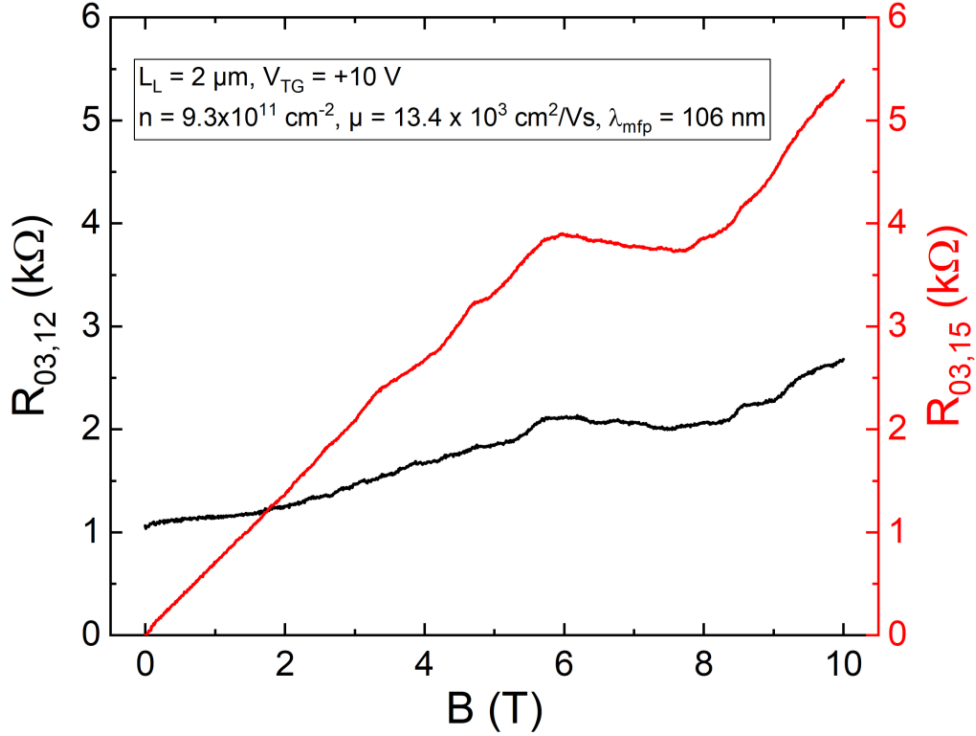

**Fig. S13.** Longitudinal resistance $R_{03,12}$ and Hall resistance $R_{03,15}$ as a function of the magnetic field at $V_{TG}$ = +10 V for the Hall bar device with $L_L$ = 2 μm from Figure 2 from the main text. $\lambda_{mfp}$ is smaller than all the contact lengths.

**Band structure calculations and maximum inverted band gap of symmetric InAs/GaInSb/InAs TQWs grown on AlSb buffer**

Band structure calculations in the main text have been performed by using the eight-band k·p Hamiltonian (24), which directly considers the interactions between $\Gamma_6$, $\Gamma_8$ and $\Gamma_7$ bands in bulk materials. In the Hamiltonian, we also consider the terms describing the strain effects arising due to mismatch of lattice constants in the buffer, QW layers and AlSb barriers. The calculations have been performed by expanding the eight-component envelope wave functions in the basis set of plane waves and by numerical solution of the eigenvalue problem. Details of calculations and the form of the Hamiltonian can be found in Ref. (*24*). Parameters for the bulk materials and valence band offsets used in the calculations are taken from Ref. (*52*).

As mentioned in the main text, higher inverted band gap values can potentially be achieved in InAs/GaInSb/InAs TQWs by increasing the In composition of the GaInSb alloy. To date, high-quality pseudomorphic growth has been achieved for InAs/Ga$_{1-x}$In$_x$Sb-based heterostructures with $x = 0.40$ (*51*). Therefore, we also further consider three-layer InAs/Ga$_{0.60}$In$_{0.40}$Sb QWs with the same In content. Even though previous theoretical studies predict the implementation of a band gap for the TI state in InAs/Ga$_{0.60}$In$_{0.40}$Sb TQWs of 60 meV grown on specific buffer (*25*), the implementation of a TQW with such a gap value is not possible due to the excessively high strain arising in the QW layers. Therefore, we further restrict ourselves to the more realistic case of InAs/Ga$_{0.60}$In$_{0.40}$Sb TQWs grown on (001) AlSb buffer. Recently, InAs/Ga$_{0.60}$In$_{0.40}$Sb QWs have been studied in BQW geometry (*38*). Figs. S14A and B summarize our calculations performed for realistic InAs/Ga$_{0.60}$In$_{0.40}$Sb TQWs. For realistic values of the layer widths, the band gap in the TI state for this TQW can reach 50 meV, which is nearly 2 times larger than the band gap of the sample studied in the main text.

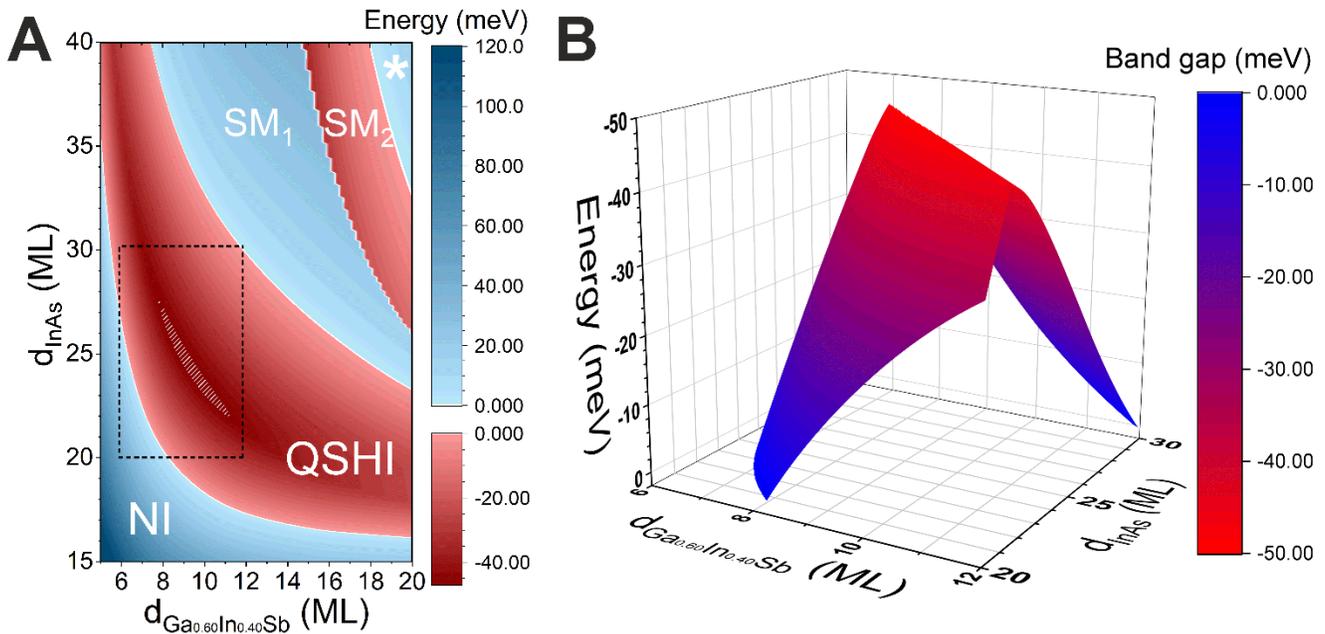

**Fig. S14.** (**A**) Colormap diagram of symmetric InAs/Ga$_{0.60}$In$_{0.40}$Sb/InAs TQWs grown on (001) AlSb buffer. NI and QSHI regions are marked, while the white region represents the area where the band gap is close to its maximum value. The regions SM$_1$ and SM$_2$ confirm the semimetals with single and double band inversions, respectively. The energy for these SM states represents the overlapping between conduction and valence bands. The blue region marked by asterisk corresponds to higher-order insulator state (*63*). Here, 1 ML corresponds to half of the lattice constant of the bulk material. (**B**) Band gap as a function of the layer widths of InAs/Ga$_{0.60}$In$_{0.40}$Sb TQWs in the rectangular region marked by dotted edges in panel (**A**).

## Hysteresis width with increasing temperatures

As the temperature increases, the hysteresis width $\varDelta V$ (difference in the peak positions for both gate voltage sweep directions) also increases. Figure S15 shows $\varDelta V$ as a function of temperature for the three devices presented in Fig. 4 of the manuscript. With increasing temperature, the up-sweep peak (used in the gate-training procedure) shifts to lower gate voltages, and the hysteresis width grows. At certain temperatures, the gate-trained up-sweep peak moves outside the accessible gate-voltage range, making gate training no longer applicable. These issues originate from the specifics of sample growth and fabrication. They might be mitigated by utilizing different capping layers and/or gate dielectrics.

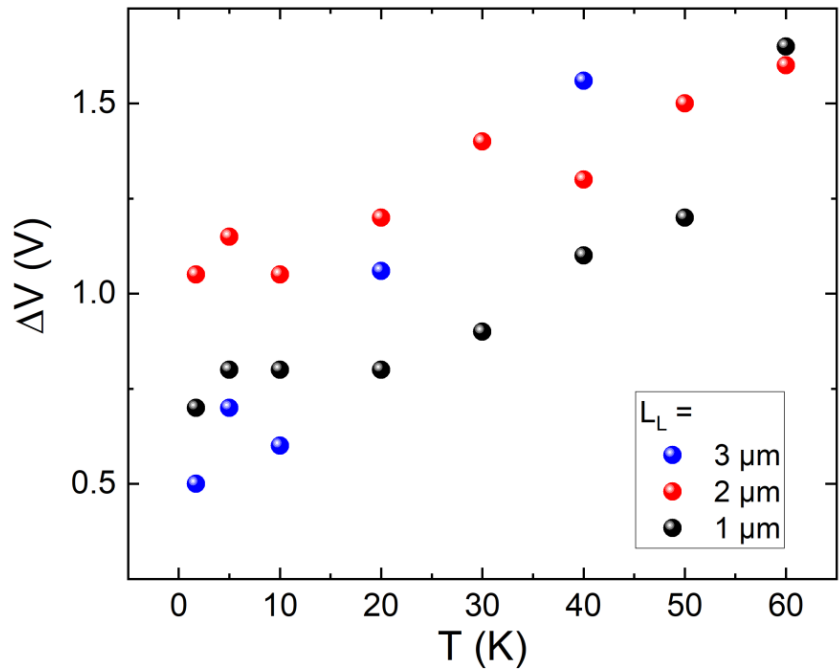

**Fig. S15.** Hysteresis width *ΔV* as a function of temperature for the three devices from Fig. 4 of the manuscript. *ΔV* increases with temperature impeding the observation of the QSHE at even higher temperatures.

In addition to hysteresis, the moderate band gap and the resulting thermally activated carriers leading to residual bulk conductivity also limit high-temperature performance.